\documentclass[preprint,preprintnumbers,endnote,nofootinbib,prd,9pt,superscriptaddress]{revtex4}

\usepackage[utf8]{inputenc}
\usepackage{graphicx}

\usepackage[format=plain,justification=raggedright, singlelinecheck=false, font=small]{caption}
\usepackage[format=plain,justification=raggedright, singlelinecheck=false,font=small]{subcaption}

\usepackage{amssymb}
\usepackage{amsxtra}
\usepackage{amsmath}
\usepackage{booktabs,multirow,tabularx}
\usepackage{slashed}
\usepackage{float}
\usepackage{placeins}
\usepackage{rotating}
\usepackage{lscape}
\usepackage{color}
\usepackage{hyperref}
\usepackage{cancel}
\usepackage[Q=yes,pverb-linebreak=no]{examplep}

\DeclareGraphicsExtensions{.pdf}
\graphicspath{{./figures/}}

\newcommand{\gsim}{\gtrsim}

\newcommand{\ord}[1]{\mathcal{O}{(#1)}}

\newcommand{\gev}{\,\textrm{GeV}}

\def\beq{\begin{equation}}
\def\bea{\begin{eqnarray}}
\def\eeq{\end{equation}}
\def\eea{\end{eqnarray}}
\def\beqal{\begin{align}}
\def\endal{\end{align}}

\makeatletter
\newcommand\footnoteref[1]{\protected@xdef\@thefnmark{\ref{#1}}\@footnotemark}
\makeatother

\DeclareFontFamily{U}{cbgreek}{}
\DeclareFontShape{U}{cbgreek}{m}{n}{
        <-6>    grmn0500
        <6-7>   grmn0600
        <7-8>   grmn0700
        <8-9>   grmn0800
        <9-10>  grmn0900
        <10-12> grmn1000
        <12-17> grmn1200
        <17->   grmn1728
      }{}
\DeclareFontShape{U}{cbgreek}{bx}{n}{
        <-6>    grxn0500
        <6-7>   grxn0600
        <7-8>   grxn0700
        <8-9>   grxn0800
        <9-10>  grxn0900
        <10-12> grxn1000
        <12-17> grxn1200
        <17->   grxn1728
      }{}

\makeatletter
\newcommand{\normalorbold}{%
  \ifnum\pdf@strcmp{\math@version}{bold}=\z@ bx\else m\fi
}
\makeatother

\begin{document}

\title{\boldmath Searching for Flavor-Violating ALPs in Higgs Decays}

\author{Hooman Davoudiasl\footnote{email: hooman@bnl.gov}}

\affiliation{High Energy Theory Group, Physics Department Brookhaven National Laboratory, Upton, NY 11973, USA}

\author{Roman Marcarelli\footnote{email: roman.marcarelli@colorado.edu}}

\author{Nicholas Miesch\footnote{email: nicholas.miesch@colorado.edu}}

\author{Ethan T. Neil\footnote{email: ethan.neil@colorado.edu}}

\affiliation{Department of Physics, University of Colorado, Boulder, Colorado 80309, USA}

\begin{abstract}
Pseudo-scalar particles, often referred to as axion-like-particles (ALPs), arise in a variety of theoretical contexts.  The phenomenology of such states is typically studied assuming flavor-conserving interactions, yet they can in principle have flavor-violating (FV) couplings to fermions.  We consider this general possibility, focusing on models where the ALP has non-negligible coupling to the Standard Model Higgs boson $h$.  For a lepton FV ALP $a$ of mass $m_a \gsim 2$~GeV, $a\to \tau \ell$, where $\ell\neq \tau$ is a charged lepton, could have $\ord{1}$ branching fraction, leading to potentially detectable $h \to a a \to \tau \ell \tau \ell$ at the LHC and its future program.  We examine this possibility, in light of existing bounds on FV processes, in a general effective theory.  We obtain constraints on the effective couplings from both prompt and long-lifetime searches at the LHC; some projections for envisioned measurements are also provided.  The implications of the recently announced first results of the muon $g-2$ measurement at Fermilab for the ALP  interactions considered in our work are also briefly discussed.

\end{abstract}

\maketitle

\section{Introduction\label{sec:intro}}

The discovery of the Higgs boson has established the Standard Model (SM) as a potentially complete description of the visible sector of the Universe.  However, robust experimental data and numerous observations have led to the firm conclusion that ingredients beyond the SM are required to explain 
neutrino flavor oscillations - and hence tiny neutrino masses - as well as the gravitationally observed, but otherwise unknown substance - {\it i.e.} dark matter (DM) - which is the dominant form of cosmic 
matter.  Currently, there are numerous proposals to explain these and some other open questions 
of particle physics and cosmology.  In particular, 
the possibility that the new physics resides in a separate  
sector and may have a rich spectrum of states and non-trivial dynamics remains a viable option.  
It is then interesting to ask what the potentially promising signals for those new phenomena are.  

A possibility that arises in various physical contexts is the appearance of pseudo-scalar particles, associated with broken global symmetries.  In the SM, spontaneous chiral symmetry breaking in the QCD sector gives rise to pions, which are the corresponding pseudo-Nambu-Goldstone bosons (PNGBs).  In general, many models of new physics 
contain such pseudo-scalars, often called axion-like particles (ALPs).  A well-known example is the QCD axion of the Peccei-Quinn mechanism \cite{Peccei:1977hh,Peccei:1977ur}, proposed to address why strong interactions 
seem to entail negligible, if any, CP violation.  Given their relative ubiquity in new physics models, ALPs are well-motivated targets for phenomenology.  We will denote a generic ALP by $a$, in what follows.

In this work, we entertain the possibility that the ALP couples to ordinary leptons with significant flavor-violating couplings.  Such interactions can naturally appear if the new-physics sector containing ALPs is linked to lepton flavor violation, including generation of neutrino masses and mixing.  The possibility of flavor-violating ALP-lepton couplings has been studied before - see for example Refs.~\cite{Berezhiani:1989fp,Cornella:2019uxs,Bauer:2019gfk,Calibbi:2020jvd,Ma:2021jkp} - but we will focus in particular on collider signals which occur when the ALP also couples to the SM Higgs boson $h$.  This opens the possibility of decays of the form $h \rightarrow aa \rightarrow (\ell \ell') (\ell \ell')$ where $\ell \neq \ell'$.  This is a distinctive final state with very low SM background.

Flavor-violating decays of the Higgs boson of this form have been studied previously in \cite{Evans:2019xer}, but in that work the intermediate particles were taken to be scalar, rather than pseudoscalar.  With pseudoscalars, the parameter space of constraints will be somewhat different.  In particular, pseudoscalar decay widths are proportional to the mass squared of the final-state particles, so decays of the form $a \rightarrow \tau \ell$ will be generally dominant for $m_a \gsim 2$~GeV.  We will study the prospects for detection of this ALP signal at the LHC, as well as envisioned experiments.

In particular, we will focus on the parameter range $2\ {\rm GeV} \lesssim m_a \lesssim m_h/2$ where the signal of interest will be strongest.  Other constraints, including constraints from astrophysical observations like SN1987a and beam-dump experiments, are not generally applicable in this range \cite{Bauer:2018uxu}.  Constraints from direct searches for lepton flavor violating (LFV) processes  such as $\tau \to 3 \, \mu$ are also weaker in this parameter range \cite{Cornella:2019uxs}, allowing for stronger lepton-ALP couplings, perhaps even sufficient to explain the anomalies in the electron and muon magnetic moments as studied in \cite{Bauer:2019gfk}. As such, we will examine flavor conserving as well as lepton flavor violating couplings necessary to explain the muon $g-2$ anomaly in our model, taking into account existing limits on LFV interactions.

For our phenomenological study, we will adopt an effective field theory (EFT) approach, valid at or below the weak scale.  Such an EFT is assumed to descend from an ultraviolet (UV) complete theory at higher scales.  By using the EFT framework, we obtain results in terms of the size of the coefficients of various operators, rendering our analysis largely model-independent.  These results can then be recast for specific models in terms of their relevant mass scales and couplings.  We will consider signatures from both prompt and displaced decays of the ALP at the LHC. 

Particles with long lifetimes necessarily have a small decay width, which corresponds to very small couplings of the particle to its decay products.  This can arise naturally for ALPs, since the interaction terms in the Lagrangian are typically suppressed by some higher energy scale. As a result, displaced decays may be the best way to search for ALPs from beyond the SM scenarios. If the particles are long-lived enough, they may escape the LHC long before decaying, so proposals for future detectors such as such as MATHUSLA~\cite{Chou:2016lxi,Curtin:2018mvb, Lubatti:2019vkf,Alpigiani:2020tva}, CODEX-b~\cite{Aielli:2019ivi}, and ANUBIS~\cite{Bauer:2019vqk} may be the best way to look for them. The combination of long-lived ALP states with flavor-violating lepton couplings is motivated by a specific UV completion~\cite{Davoudiasl:2017zws} in which a dark sector gives rise to both neutrino masses and dark matter.

The paper is organized as follows.  In Section \ref{sec:ET}, we describe the effective theory to be used in our collider studies.  Section \ref{sec:FV} considers bounds on the parameter space of the effective theory due to LFV constraints from lepton conversion, and Section \ref{sec:Hdecay} considers exclusions and projected exclusions on the Higgs branching fraction into ALPs for prompt and displaced decays. In Section \ref{sec:g-2}, we examine the region of our parameter space necessary to explain the muon $g-2$ anomaly, finding that in such a scenario LFV couplings of the ALP must be greatly suppressed compared to lepton-flavor conserving (LFC) couplings.

\section{Effective Theory\label{sec:ET}}

There has been substantial work in the literature on constructing a low-energy effective theory for ALPs coupled to the SM  \cite{Georgi:1986df,Chang:2000ii,Marciano:2016yhf,Bauer:2017ris,Brivio:2017ije,Bauer:2018uxu,Cornella:2019uxs,Bauer:2020jbp,Chala:2020wvs,Calibbi:2020jvd,Ma:2021jkp,Buen-Abad:2021fwq}.  Here, we focus on a subset of all possible couplings which are most relevant for the signal of interest.  We will adopt the following simplified effective Lagrangian in the electroweak-broken phase:

\beq
\mathcal{L} = \frac{1}{2} (\partial_\mu a)^2 - \frac{1}{2} m_a^2 a^2 + \mathcal{L}_{\ell} + \mathcal{L}_g + \mathcal{L}_{h}\label{eq:alp_L}
\eeq
where
\begin{align} 
\mathcal{L}_{\ell} &= \partial_\mu a \sum_{\ell, \ell'} \left[ \frac{C_{\ell \ell'}}{\Lambda} \bar{\ell} \gamma^\mu \gamma_5 \ell' + \textrm{h.c.} \right]\label{eq:Ll}, \\
\mathcal{L}_{g} &= 4\pi \alpha \frac{C_{\gamma \gamma}}{\Lambda} a F_{\mu \nu} \tilde{F}^{\mu \nu} + ... \\
\mathcal{L}_{h} &= \frac{C_{ah}}{\Lambda^2} v (\partial_\mu a)^2 h + \frac{C_{ah}'}{\Lambda^2} v m_a^2 a^2 h + ...
\end{align}

We assume that the other neglected couplings, {\it e.g.} direct coupling of the ALP to light quarks, are negligible.  We include only parity-conserving couplings, neglecting the possibility of the lepton vector current coupling to the ALP. In our main phenomenological study, we will assume that all of the leptonic couplings $C_{\ell \ell'}$, both flavor-diagonal and off-diagonal, are roughly equal in magnitude.  The possibility of a direct coupling of $a$ to neutrinos is also omitted here, see Ref.~\cite{Bauer:2020jbp} for a discussion of including such a term; in the present case such a coupling would not have any direct impact on the phenomenology, as the decay $a \rightarrow \nu \bar{\nu}$ is proportional to $m_\nu^2$ or induced by a high-dimension operator and therefore extremely suppressed \cite{Bauer:2017ris}.

We will consider the scenario where the tree-level coupling $C_{\gamma\gamma} = 0$, but we note that such a coupling can still be generated at loop order. The most general decay width for $a\rightarrow \gamma\gamma$ is given by \cite{Bauer:2017ris}
\beq
    \Gamma(a\rightarrow \gamma\gamma) = \frac{4\pi\alpha^2m_a^3}{\Lambda^2}|C_{\gamma\gamma}^{\rm eff}|^2
\eeq
where the coupling $C_{\gamma \gamma}^{\rm eff}$ is

\beq
C_{\gamma \gamma}^{\rm eff} = C_{\gamma \gamma} + \sum_f \frac{N_c^f Q_f^2}{8\pi^2} C_{ff} B_1\left(\frac{4m_f^2}{m_a^2} \right). \label{eq:Cggeff}
\eeq
(Note that our definition of $C_{ff}$ differs by a factor of 2 from $c_{ff}$ in the reference.) Here, $B_1(\tau) = 1 - \tau f(\tau)^2$, where 
\begin{align}
    f(\tau) &= \begin{cases}\sin^{-1}\left(\frac{1}{\sqrt{\tau}}\right) & \tau \geq 1,\\
    \frac{\pi}{2} + \frac{i}{2}\log{\frac{1 + \sqrt{1 - \tau}}{1 - \sqrt{1-\tau}}}, & \tau < 1\end{cases}
\end{align}
It follows that $B_1(4m_f^2/m_a^2)\approx 1$ if $m_f \ll m_a$.   Assuming that only the lepton couplings are non-negligible gives $Q_f^2 = 1$ and $N_c^f = 1$. Hence, there will still be some branching fraction of $a$ into photons even with $C_{\gamma\gamma} = 0$.  In the mass range of interest, this branching fraction is small compared to those of leptonic decay modes, as we will now demonstrate.

The decay width of the ALP into leptons $\ell$,$\ell'$ is given by
\begin{align}
\Gamma(a\rightarrow {\ell \ell'}) &= \frac{|C_{\ell\ell'}|^2}{8\pi\Lambda^2}\frac{(m_{\ell'}+m_{\ell})^2}{m_a^3}(m_a^2-(m_{\ell'}-m_{\ell})^2)\nonumber\\
    &\ \ \ \ \ \ \ \ \times\sqrt{[m_a^2 - (m_{\ell'} - m_{\ell})^2][m_a^2 - (m_{\ell'} + m_{\ell})^2]}\\
    &\approx \frac{|C_{\ell\ell'}|^2}{8\pi \Lambda^2}\frac{m_{\ell'}^2}{m_a^3}(m_a^2-m_{\ell'}^2)^2 \label{eq:all_decay}
\end{align}
assuming that $m_\ell \ll m_{\ell'}$ on the last line.  Note that the decay width is roughly proportional to the square of the heavier daughter-lepton mass, so that even if there is no hierarchy among the couplings $|C_{\ell \ell'}|$, decays involving $\tau$ leptons will generally be dominant.  This is one of the key differences between ALP and scalar decays into pairs of fermions.  Whereas scalars can couple to fermions with arbitrary couplings, the origin of ALPs as Goldstone bosons mandates that they interact through derivatives (by shift-symmetry) as in Eq.~(\ref{eq:Ll}), which leads to couplings proportional to masses using the equations of motion.  

If the photon coupling is only generated at loop order, then for $m_a \gtrsim m_\tau$, we have
\begin{align}
    \frac{\Gamma(a\rightarrow \gamma\gamma)}{\Gamma(a\rightarrow \tau\ell)}\approx \frac{\frac{4\pi\alpha^2m_a^3}{\Lambda^2}\left(\frac{1}{8\pi^2}|C_{\tau\ell}|^2\right)^2}{\frac{|C_{\tau\ell}|^2}{8\pi \Lambda^2}m_{\tau}^2m_a} \approx \frac{m_a^2}{(600m_\tau)^2}
\end{align}
where we have used $C_{\gamma\gamma}^{\text{eff}}= \mathcal{O}\left(\frac{1}{8\pi^2}C_{\tau\ell}\right) \approx 0.01C_{\tau \ell}$. It follows that the photon branching fraction only becomes important when $m_a \gtrsim 1200\,\rm{GeV}$, far outside the range of masses considered. Although this argument is only technically valid for $m_a \gg 2\,\rm{GeV}$, a numerical check shows that $\mathcal{B}(a\rightarrow\gamma\gamma) \sim 10^{-7}$ even for $m_a = 2\,\rm{GeV}$, because the subdominant leptonic modes such as $a\rightarrow \mu\mu$ still have a much larger width than the width into photons. Hence, to very good approximation the ALP in our model decays only through leptonic modes, and mostly to modes with at least one $\tau$, so that is what we will focus on for the remainder of the study.

As previously noted, in the following we restrict the model parameter space to 
\beq
2\ {\rm GeV} \lesssim m_a \lesssim m_h/2.
\eeq
This ensures that the key process of interest $h \rightarrow aa \rightarrow (\ell \ell')(\ell \ell')$ can proceed fully on-shell to all three lepton generations.  We will also primarily consider the case that the leptonic couplings are universal (e.g., $C_{\ell\ell'} \approx \text{const.}$), and we hereby refer to this single constant as $\overline{C}_{LL}$. Additional phenomenology can be done by considering various hierarchies of the different couplings, and we consider such a scenario in more detail in Section \ref{sec:g-2}. Due to the natural hierarchy which arises from the lepton mass dependence of the ALP-lepton decays, one would need to introduce an opposing hierarchy in the couplings $C_{\ell \ell'}$ in order to significantly alter our phenomenological results.

\section{Constraints on flavor-violating couplings from lepton conversion\label{sec:FV}}

Before we proceed to collider phenomenology of Higgs decays, we first consider existing experimental constraints on the flavor-violating lepton couplings in our model.  Such constraints have been studied previously for ALPs in Ref.~\cite{Cornella:2019uxs}.  In the region of interest $m_a \gtrsim m_\tau$ that we have identified, the single strongest constraint on the couplings $C_{\ell \ell'}$ currently comes from searches for lepton conversion of the form $\ell \rightarrow \ell' \gamma$.  We will thus focus on this particular signal as a constraint.  Future improvements in searches for decays of the form $\mu \rightarrow 3e$, $\mu \rightarrow e \gamma \gamma$ and $\tau \rightarrow 3(e,\mu)$ may lead to relevant constraints even for $m_a \gtrsim m_\tau$; see Ref.~\cite{Cornella:2019uxs} for details.

\begin{figure}[t!]
    \centering
    \includegraphics[width = 0.8\linewidth]{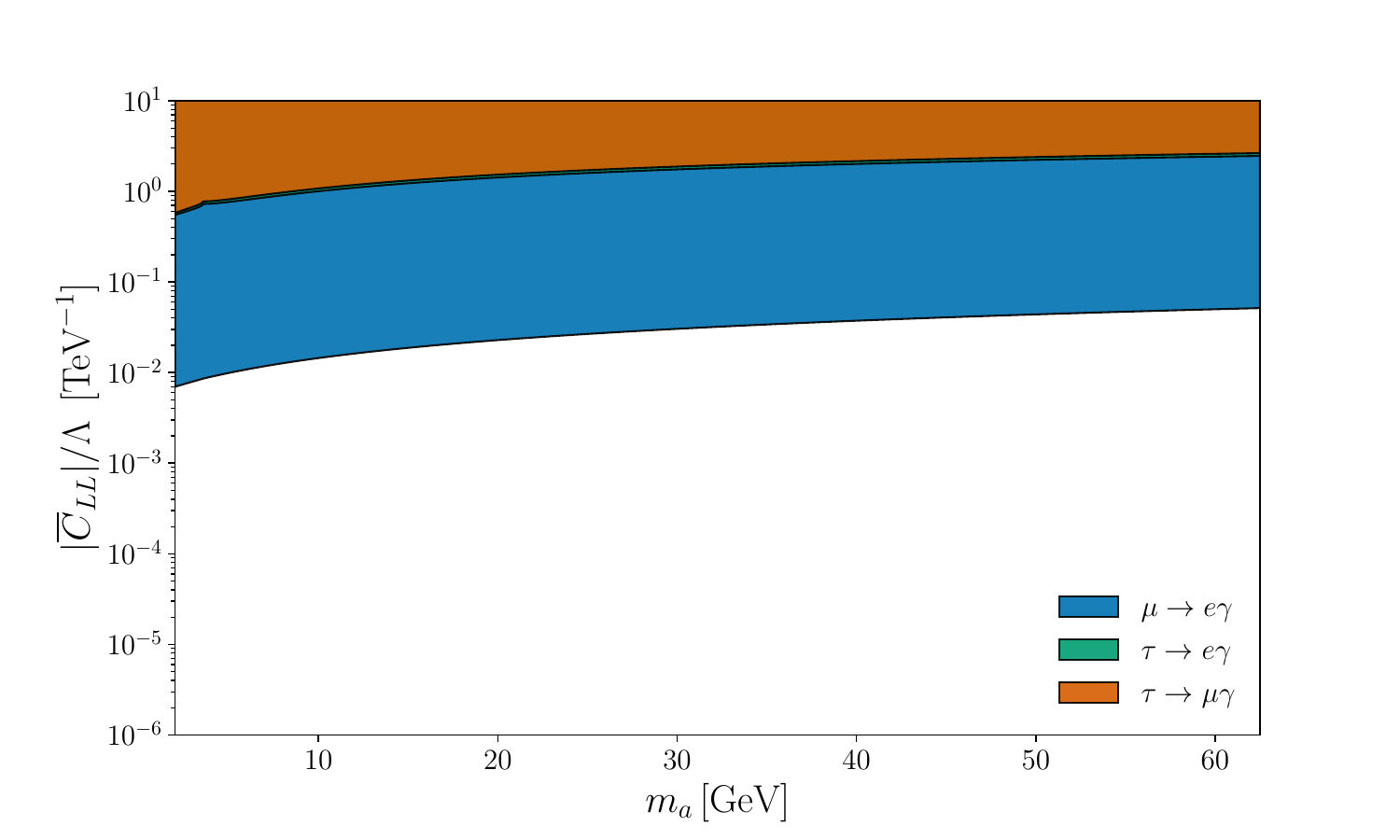}
    \caption{Constraints on the coupling $\overline{C}_{LL}/\Lambda$ from  ($\mu \rightarrow e\gamma$)\cite{TheMEG:2016wtm} and ($\tau\rightarrow \ell\gamma$)\cite{Aubert:2009ag}. The most stringent constraint is given by the process $\mu\rightarrow e\gamma$, so this is the LFV constraint we will adopt for the rest of our analysis.}
    \label{fig:LFV}
\end{figure}

From Ref.~\cite{Cornella:2019uxs}, the decay rate for $\ell\rightarrow \ell'\gamma$  is given by
\beq
\Gamma(\ell\rightarrow \ell'\gamma) = \frac{\alpha m_{\ell}}{2}|\mathcal{F}_2^{\ell\ell'}(0)|^2\label{eq:leptconv},
\eeq
where the form factor $\mathcal{F}_2^{\ell\ell'}(0)$ is defined in Appendix~\ref{sec:lfv_fns}.  We use $C_{\gamma\gamma}^{\text{eff}}$ for the photon coupling in Eq.~(\ref{eq:F2lin}). To be more precise, one should perform the explicit two-loop calculation to consider the effect of the loop-induced photon coupling, since the ALP and a photon in the loop can be off-shell, but this is beyond the scope of our analysis. 

Under the previously stated assumption that $C_{\ell\ell'}\approx \overline{C}_{LL}$, $\mathcal{F}_2^{\ell\ell'}(0)$ is proportional to $|\overline{C}_{LL}|^2$, so we can define $\mathcal{F}_2^{\ell\ell'}(0) \equiv |\overline{C}_{LL}|^2 \mathcal{F}^{\ell\ell'}_{2,0}$, where $\mathcal{F}^{\ell\ell'}_{2,0}$ is the  form factor evaluated at unit $C_{\ell\ell'}$. With this definition, we obtain 
\beq
    \Gamma(\ell\rightarrow \ell'\gamma) = \frac{\alpha m_{\ell}}{2}|\overline{C}_{LL}|^4\left|\mathcal{F}_{2,0}^{\ell\ell'}\right|^2.
\eeq
Hence, constraints on the decay rate can be cast to constraints on $|\overline{C}_{LL}|/\Lambda$ via
\beq
    |\overline{C}_{LL}|\leq \left(\frac{2\Gamma(\ell\rightarrow \ell'\gamma)}{\alpha m_{\ell}\left|\mathcal{F}_{2,0}^{\ell\ell'}\right|^2}\right)^{1/4}.
\eeq
Constraints on $\Gamma(\mu\rightarrow e\gamma)$ are obtained from the MEG \cite{TheMEG:2016wtm} experiment, whereas $\Gamma(\tau\rightarrow\ell\gamma)$ constraints are obtained from BaBar \cite{Aubert:2009ag}. In particular, at 90\% CL,
\bea
    \Gamma(\mu\rightarrow e\gamma) &< 1.3\times 10^{-31}\,\rm{GeV},\\
    \Gamma(\tau \rightarrow e\gamma) &<7.6\times 10^{-20}\,\rm{GeV},\\
    \Gamma(\tau \rightarrow \mu\gamma) &< 1.0\times 10^{-19}\,\rm{GeV}.
\eea
The resulting exclusion plots for the mass range of interest ($2\text{ GeV}\lesssim m_a \lesssim m_h/2$) are shown in Fig.~\ref{fig:LFV}. The most stringent constraint comes from the bound on $\Gamma(\mu\rightarrow e\gamma)$.

\section{Higgs decay signal\label{sec:Hdecay}}

Due to the coupling of the ALP to leptons, one might expect that limits could be obtained from electron-positron colliders such as LEP and CESR, since such a particle could be produced via $e^+ e^- \rightarrow a$. However, ALP production from $e^+ e^-$ is severely suppressed by $m_e^2/\Lambda^2$, due to the derivative coupling of the ALP. Since the ALP considered also does not couple to quarks, we do not expect there to be direct production from hadron colliders such as the LHC. Hence, we expect that Higgs decays are the best way to look for such a particle. For most of the mass region considered, over $99\%$ of ALPs decay via $a\rightarrow \tau \ell$, so the best constraints will come from the decay mode $h \rightarrow aa \rightarrow (\tau \ell)(\tau \ell')$ . 

The ALP coupling $C_{\ell \ell'}$ can also generate decays of the form $Z \rightarrow (\ell \ell') a \rightarrow (\ell \ell') (\ell \ell')$, but as was shown in Ref.~\cite{Evans:2019xer}, the limits from $Z$ decays are much less significant than those arising from Higgs decays (this was shown for a scalar boson in the reference, but we determined the differences for a pseudoscalar are minor, with corrections proportional to $\frac{m_\ell}{m_Z}$.) For an ALP that only couples to the leptons through a derivative, the $Z$ decay is also even further suppressed by a factor of the lepton mass.  Therefore, we will safely ignore that process for the remainder of this paper.

The Higgs decay rate into ALPs is given by:
\beq
\Gamma({h\rightarrow aa}) = \frac{1}{32\pi} \frac{v^2 m_h^3}{\Lambda^4} \sqrt{1 - \frac{4m_a^2}{m_h^2}} \left(C_{ah} - 2(C_{ah} + C_{ah}') \frac{m_a^2}{m_h^2}\right)^2
\eeq
Defining the quantities
\beq
\overline{C}_{ah} \equiv C_{ah} - \frac{2m_a^2}{m_h^2-2m_a^2}C_{ah}^{\prime} 
\label{eq:Cahbar}
\eeq
and 
\begin{align}
    \Gamma_0(h\rightarrow aa) = \frac{1}{32\pi}\frac{v^2 m_h^3}{\Lambda^4}\sqrt{1-\frac{4m_a^2}{m_h^2}},
\end{align}this decay rate can be written as
\beq
    \Gamma({h\rightarrow aa}) = |\overline{C}_{ah}(m_a)|^2 \Gamma_{0}(h\rightarrow aa).
\eeq
It will be interesting to place constraints on the parameters $C_{ah}$ and $C_{ah}'$ when given constraints on the branching fraction of a certain process $\mathcal{B}(h\rightarrow 2a\rightarrow X)$. This branching fraction can be written as
\begin{align}
    \mathcal{B}(h\rightarrow 2a\rightarrow X) &= \frac{\Gamma(h\rightarrow aa)}{\Gamma(h\rightarrow aa) + \Gamma_{h,\text{sm}}}\mathcal{B}(2a\rightarrow X)\\
    &= \frac{|\overline{C}_{ah}|^2\Gamma_{0}(h\rightarrow aa)}{|\overline{C}_{ah}|^2\Gamma_{0}(h\rightarrow aa) + \Gamma_{h,\text{sm}}}\mathcal{B}(2a\rightarrow X)
\end{align}
which is a monotonic function in $|\overline{C}_{ah}|^2$. Hence, if there is a constraint $\mathcal{B}\leq \mathcal{B}_{\text{max}}$, this translates to a constraint
\beq
|\overline{C}_{ah}|^2 \leq \frac{\mathcal{B}_{\text{max}}}{\mathcal{B}(2a\rightarrow X)-\mathcal{B}_{\text{max}}}\frac{\Gamma_{h,\text{sm}}}{\Gamma_{0}(h\rightarrow aa)}\equiv (\overline{C}_{ah,\text{max}})^2.\label{eq:Cah_eff_constraint}
\eeq
From here, an allowed region can be found for $C_{ah}$ in terms of $C_{ah}'$ or vice versa. In particular, 
\beq
\left|C_{ah} - \frac{2m_a^2}{m_h^2 - 2m_a^2}C_{ah}'\right| \leq \overline{C}_{ah,\text{max}}.
\eeq
Hence, it is most natural to place a direct constraint on $\overline{C}_{ah}$, and then translate the result to specific values of $C_{ah}$ or $C_{ah}'$ as desired.

\subsection{Constraints from existing searches}
One feature of our model which makes it difficult to constrain is the dependence of the ALP coupling on the masses of the final-state leptons.  Since $m_\tau\gg m_e,m_\mu$, the ALP will decay with at least one $\tau$ in the final state $\mathcal{O}(\tfrac{8m_\tau^2 }{8 m_\tau^2 + 6m_\mu^2})\approx 99.7\%$ of the time. Hence, any study which attempts to reconstruct pairs of electrons and muons is not very constraining for our model, since there is likely to be missing energy associated with such pairs.

\begin{figure}
    \centering
    \includegraphics[width=0.9\linewidth]{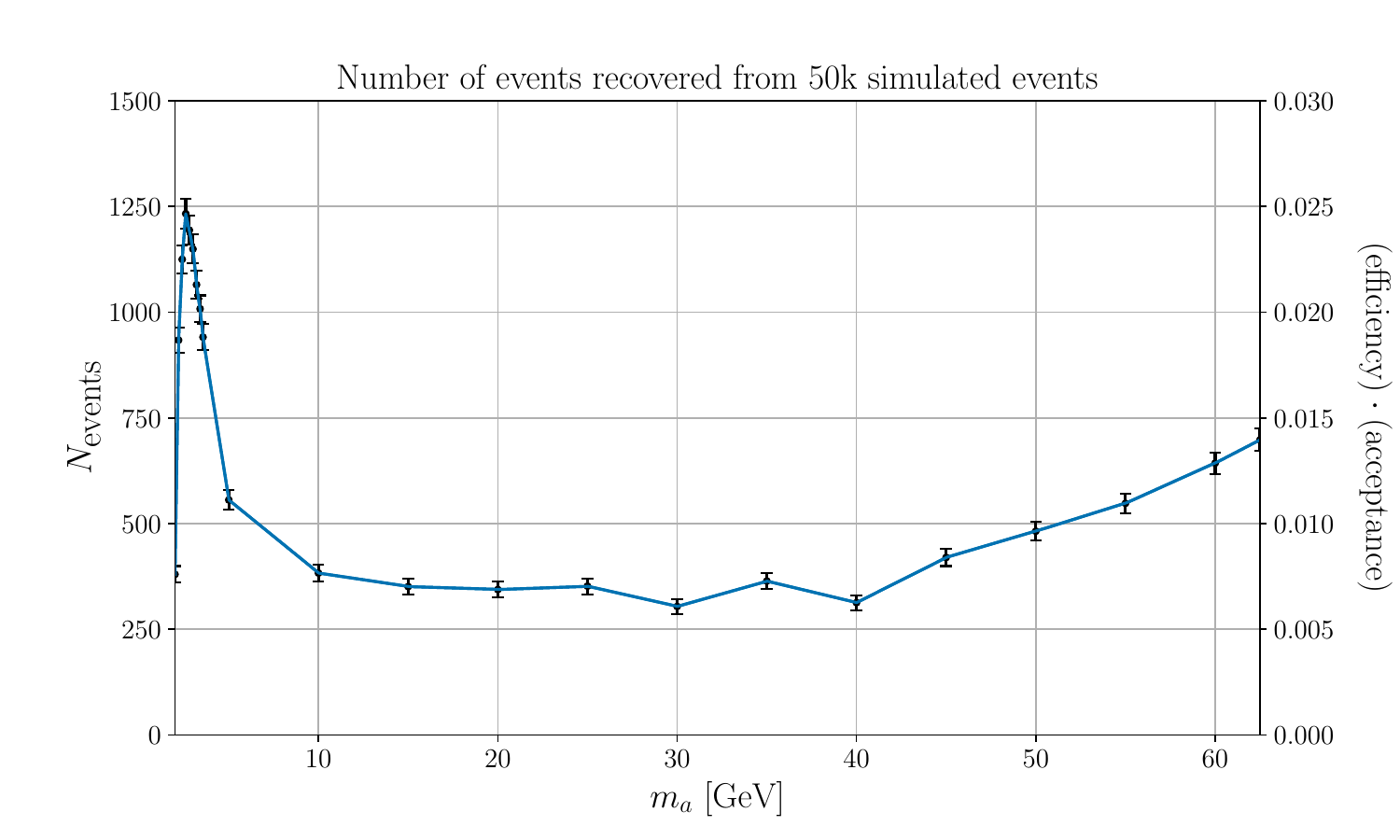}
    \caption{The number of OSSF0 events recovered following the analysis of Ref.~\cite{Sirunyan:2019bgz}, from 50k simulated events. Uncertainties were calculated using Poisson statistics.}
    \label{fig:efficiency}
\end{figure}

\begin{figure}[t!]
    \centering
    \includegraphics[width=0.8\linewidth]{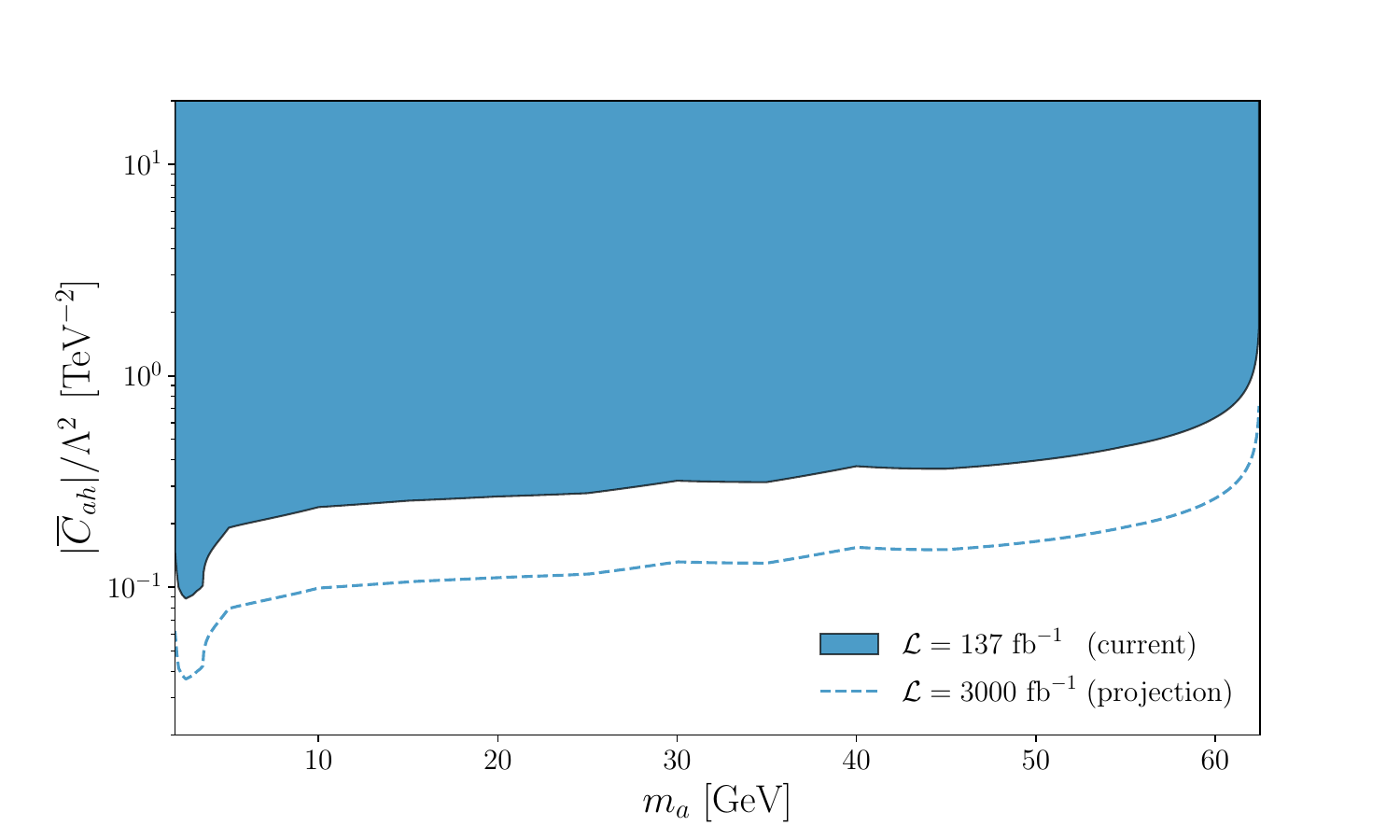}
    \caption{Constraints on $\overline{C}_{ah}$ for $m_a\in(2m_\tau, m_h/2)$. The blue region is the current constraint from Ref.~\cite{Sirunyan:2019bgz}, and the dashed line is the projection for $\mathcal{L} = 3000\text{ fb}^{-1}$. These limits assume the decay of the $a$ is prompt, and will be weakened for long-lived ALPs, as shown in Fig.~\ref{fig:prompt_Cah_vs_ma_lifetimes} below.}
    \label{fig:prompt_Cah_vs_ma}
\end{figure}

Here, we present constraints based on a search in Ref.~\cite{Sirunyan:2019bgz} for proton-proton collisions which result in 4 lepton final states. In particular, we are interested in the ``OSSF0'' signal, which corresponds to the events in which the number of opposite-sign, same-flavor (OSSF) lepton pairs is zero. The dominant modes that would contribute to this signal in our model are still those which contain at least one $\tau$, which then decays into an electron or muon. To this end, we simulate Higgs production via gluon-gluon fusion, with the Higgs decaying via $h\rightarrow aa \rightarrow \tau_{\ell}^{\pm}\tau_{\ell}^{\pm}\tau_{\ell'}^{\mp}\tau_{\ell'}^{\mp}$, $h\rightarrow aa \rightarrow \tau_{\ell}^{\pm}\tau_{\ell}^{\pm}\tau_{\ell'}^{\mp}\ell'^{\mp}$, or $h\rightarrow aa \rightarrow \tau_{\ell}^{\pm}\tau_{\ell}^{\pm}{\ell'}^{\mp}{\ell'}^{\mp}$.  Here, $\tau_\ell$ indicates a $\tau$ lepton which decays as $\tau \rightarrow \ell \bar{\nu} \nu$.  We denote these processes collectively as $h\rightarrow aa \rightarrow \text{OSSF0}$. The model is created using \verb|FeynRules|~\cite{Alloul:2013bka}, and the parton-level events are created in \verb|MadGraph5_aMC@NLO|~\cite{Alwall:2014hca}. We then use \verb|Pythia8|~\cite{Sjostrand:2007gs} for showering, and simulate the CMS detector with \verb|Delphes|~\cite{deFavereau:2013fsa}. We produce 50k events of the form $g g \rightarrow h, h \rightarrow a a \rightarrow \text{OSSF0}$ for the masses $m_a = 2\,\rm{GeV}$ to $m_a = 3.6\,\rm{GeV}$ with a step of $0.2\,\rm{GeV}$, and $m_a = 5\,\rm{GeV}$ to $m_a = 60\,\rm{GeV}$ with a step of $5\,\rm{GeV}$. We additionally consider $m_a = 62.5\,\rm{GeV}$ for a total of 22 different masses. The number of events recovered from the initial 50k based on the analysis in Ref.~\cite{Sirunyan:2019bgz} and the corresponding $(\text{efficiency})\cdot(\text{acceptance})$ is shown in Fig.~\ref{fig:efficiency}.
In the analysis of Ref.~\cite{Sirunyan:2019bgz}, the number of events with OSSF0 pairs predicted and observed at a luminosity of $\mathcal{L} = 137\ \text{fb}^{-1}$ is 7. Following the analysis from Ref.~\cite{Feldman:1997qc}, the 95\% confidence interval on the mean number of signal events when the number of background events and number of observed events are both 7 is (0, 6.81). A 95\% CL upper-limit on the branching fraction is then derived via
\beq
    \epsilon \sigma_{ggh}\mathcal{L}\mathcal{B}(h\rightarrow aa \rightarrow\text{OSSF0})\leq N_{\text{max}} = 6.81
\eeq
where $\epsilon = (\rm{efficiency})\cdot(\rm{acceptance})$. This constraint is then recast to a constraint on $\overline{C}_{ah}/\Lambda^2$ from Eq. (\ref{eq:Cah_eff_constraint}) by using the theoretical branching fraction for $\mathcal{B}(h\rightarrow aa\rightarrow \text{OSSF0})$ and interpolating $\epsilon$ from Fig.~\ref{fig:efficiency} to all values of mass. The results of this constraint for $\mathcal{L} = 137\text{ fb}^{-1}$ and a projected value of $\mathcal{L} = 3000\text{ fb}^{-1}$ are shown in Fig.~\ref{fig:prompt_Cah_vs_ma}. For the projection, we assume that the number of observed events is the same as the number of expected background events, which would be $(3000/137)\times 7 \approx 153$. Again following the analysis from Ref.~\cite{Feldman:1997qc}, this corresponds to a $95\%$ confidence interval on the mean number of signal events of $(0,25.9)$.

\subsection{Limits from displaced decays}\label{sec:displaced}

Since the branching fraction of the ALP into leptonic final states is nearly $100\%$, it is difficult to place a constraint on the parameters $|C_{\ell\ell'}|$ from prompt decays alone. For a certain regime of lepton couplings, the decay of the ALP will be displaced, and will thereby be subject to experiments which search for long-lived particles. Such a scenario is more natural in the ALP case than the scalar case, since there is an additional suppression of $\mathcal{O}(m_{\tau}^2/\Lambda^2)$ of the ALP decay rate, which translates to a $\sim\Lambda^2/m_\tau^2$ enhancement of the ALP lifetime.  In our model, the lifetime $\tau_a$ of the ALP is given by
\begin{align}
    \tau_a^{-1} = \sum_{\ell,\ell'}\Gamma(a\rightarrow {\ell\ell'}) = |\overline{C}_{LL}|^2\sum_{\ell,\ell'}\Gamma_{0,\ell\ell'}, \label{eq:lifetime}
\end{align}
where we have once again made the assumption that the couplings are roughly universal and given by a value $C_{\ell\ell'} = \overline{C}_{LL}$, and defined $\Gamma_{0,\ell\ell'} \equiv \Gamma(a\rightarrow{\ell\ell'})/|\overline{C}_{LL}|^2.$ 

\begin{figure}
    \includegraphics[width=0.8\linewidth]{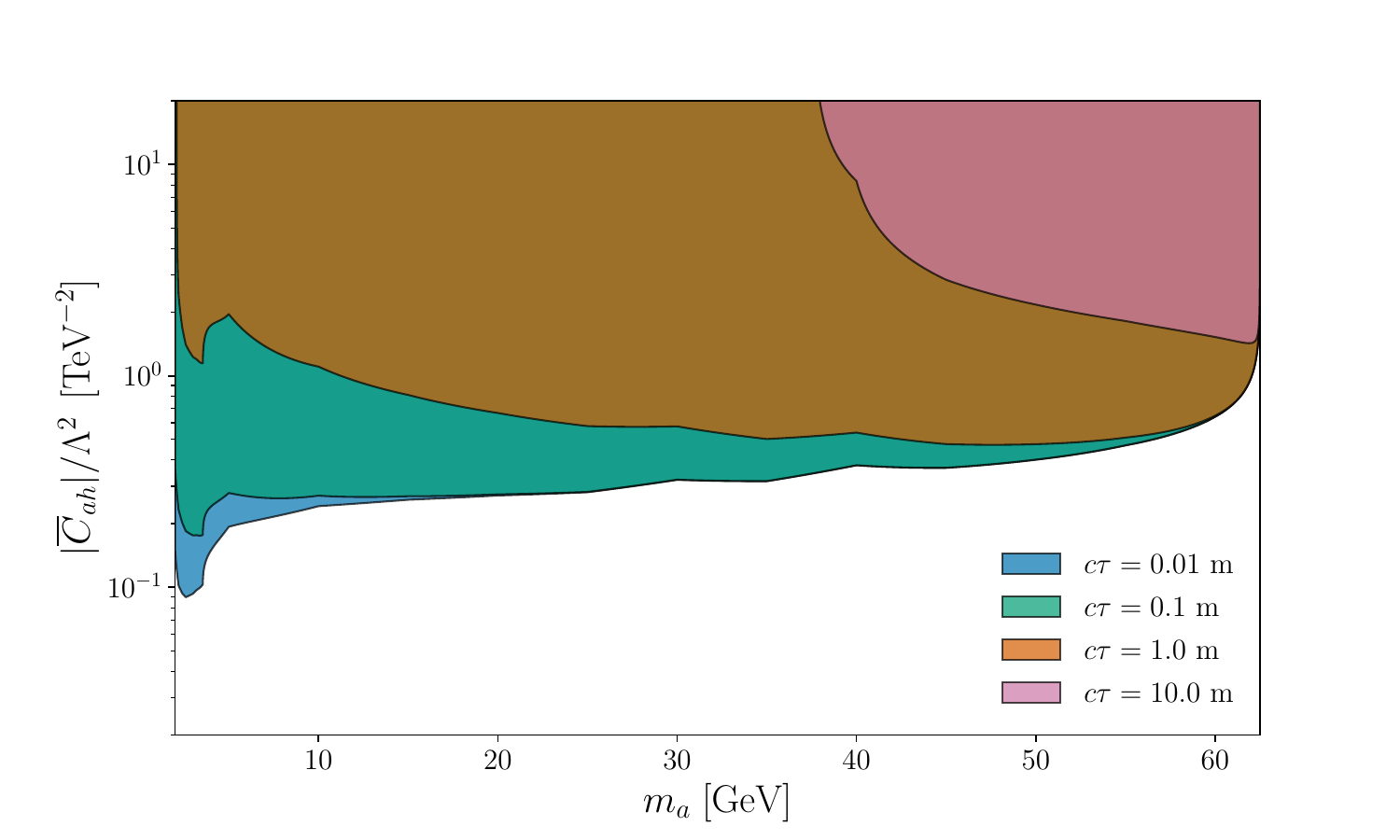}
    \caption{Constraints on $\overline{C}_{ah}$ for $m_a\in(2m_\tau, m_h/2)$ for various lifetimes of the ALP. As the lifetime increases, more ALPs escape the detector and the constraints weaken.}
    \label{fig:prompt_Cah_vs_ma_lifetimes}
\end{figure}

In the case of a long-lived ALP, the constraints in the previous section have to be modified to accommodate for the fact that the ALP may leave the detector. A formula for the fraction of pair-produced long-lived ALPs which decay in a cylindrical detector is derived in Ref.~\cite{Bauer:2017ris} to be
\beq
    f_{aa}(L_a) = \int_0^{\pi/2}d\theta\,\sin{\theta}\left(1 - e^{-L_\text{det}/[L_{a}\sin{\theta}]}\right)^2
\eeq
 where $L_{\rm{det}}$ is the radius of the detector, $\theta$ is the angle of the ALP pair w.r.t. the beam axis in the rest frame of the Higgs, and $L_a \equiv (\beta_a\gamma_a)c\tau_a$ is the decay length of the ALP in the lab frame. Here $c$ is the speed of light and we take $\gamma_a = m_h/2m_a$. Taking $L_{\text{det}} = 1.1\text{ meters}$ for the inner detector of CMS, we can recast the constraints from the previous section for a long lifetime by multiplying $\epsilon$ by the fraction $f_{aa}$. These results are demonstrated in Fig.~\ref{fig:prompt_Cah_vs_ma_lifetimes}.

In addition to recasting prompt constraints, we can also examine constraints from searches for long-lived particles at CERN. One difficulty with such a task is that the main decay mode of the ALP has a very high hadronic background (for example, from $B$-mesons) for displaced decays in the ATLAS or CMS inner detectors. As a result, we expect displaced decay analyses to become more important and fruitful when the ALP can travel into the muon spectrometer. This corresponds to $\tau_a\gsim 1\text{ m}$ or $|\bar{C}_{LL}|/\Lambda \lesssim 10^{-5}~\rm{TeV}^{-1}$ for $m_a = 10\text{ GeV}$. One work that considers such a scenario is Ref.~\cite{Aad:2019xav}, which places constraints on the Higgs decaying into long-lived particles which have displaced jets in the final state by combining data from the ATLAS inner detector with data from the muon spectrometer. In particular, they consider the scenario in which one displaced jet is found in the inner detector and the other is found in the muon spectrometer. Since most of the ALPs in our model decay with at least one $\tau$, and most $\tau$s decay hadronically, this study can be used to constrain our model.

To adapt the work for our purposes, we digitize their results for a 95\% CL upper-limit on the branching fraction for Higgs into two scalars, then use a piecewise polynomial fit to generalize it to a variety of masses and lifetimes. The plots we digitized cover an extensive region of lifetimes, but are only presented for four masses: $m_a = 10,\, 25,\, 40, $ and $55\,\rm{GeV}$. Due to this limited data along the mass axis, we account for overfitting by interpolating plots for $m_a = 15,\, 32.5, $ and $47.5\,\rm{GeV}$ using the geometric mean of the branching ratio limits corresponding to the nearest two masses, and then use these values along with the original masses to ensure that the fit function has a relatively smooth dependence along the mass axis. The fit function agrees well with the data in regions of overlap, but we stress that results are likely less accurate for $m_a \lesssim 10\,\rm{GeV}$ and $m_a\gtrsim 55\,\rm{GeV}$. For more accurate data in these mass regimes, the analysis of Ref.~\cite{Aad:2019xav} would need to be repeated. The details of the fitting procedure can be found in the Github repository Ref.~\cite{MarcarelliGithub2021}. Once again, these branching fraction constraints can be recast to constraints on $\overline{C}_{ah}/\Lambda^2$. Using Eq.\,(\ref{eq:lifetime}), we can represent the constraints in terms of $|\overline{C}_{LL}|$ to create an exclusion for $\overline{C}_{ah}/\Lambda^2$ and $\overline{C}_{LL}/\Lambda$. 

\begin{figure}
    \centering
    \includegraphics[width=0.8\linewidth]{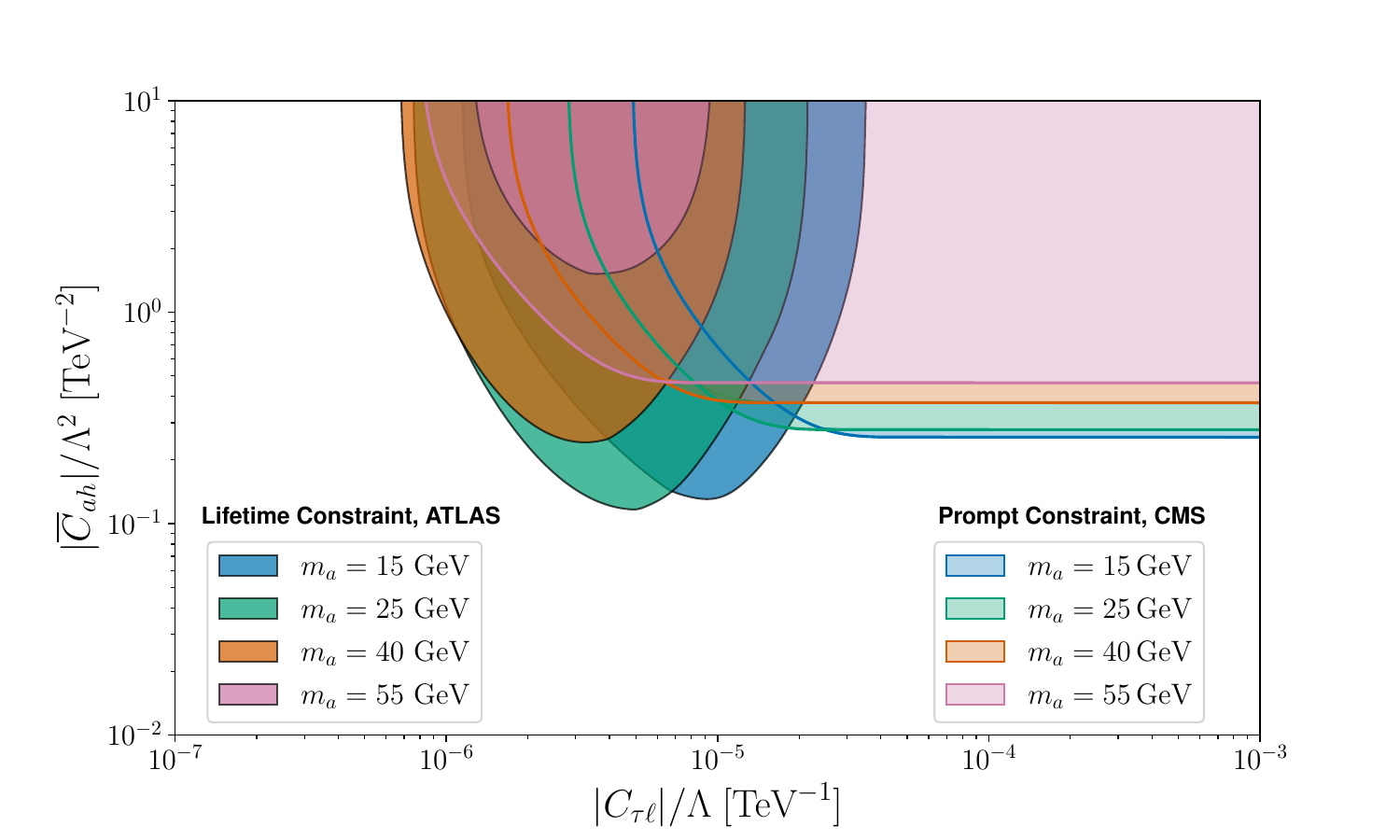}
    \caption{Exclusion plots for $C_{ah}$ vs. $C_{\tau\ell}$ at a 95\% confidence level, based on the studies Refs.~ \cite{Sirunyan:2019bgz} and \cite{Aad:2019xav}. The lifetime constraints from displaced jets~\cite{Aad:2019xav} are the solid regions, while the prompt constraints from the OSSF0 signal~\cite{Sirunyan:2019bgz} are the transparent regions.  Note that although in the model we assume a universal lepton coupling $\overline{C}_{LL}$, the exclusion here is essentially on the subset of the couplings $C_{\tau \ell}$.}
    \label{fig:lifetime_prompt_Cah_vs_Cll}
\end{figure}

These results, combined with the results from Fig.\,\ref{fig:prompt_Cah_vs_ma_lifetimes}, are shown in Fig.\,\ref{fig:lifetime_prompt_Cah_vs_Cll}. The constraints from Ref.~\cite{Aad:2019xav} are the solid regions, while the prompt constraints from Ref.~\cite{Sirunyan:2019bgz} are the transparent regions.
Based on existing searches, the model is not constrained at all for $|\overline{C}_{ah}|/\Lambda^2 \lesssim 0.1\text{ TeV}^{-1}$ or $\overline{C}_{LL}/\Lambda \lesssim 10^{-6}\text{ TeV}^{-1}$. Assuming $\overline{C}_{ah}$ and $\overline{C}_{LL}$ are $\mathcal{O}(1)$, such a model would still be completely unconstrained for $\Lambda \gtrsim 3\text{ TeV}$. At a projected luminosity of $\mathcal{L} = 3000\text{ fb}^{-1}$, the $C_{ah}$ constraints can be improved by a factor of $\mathcal{O}(5$-$10)$, but $\overline{C}_{LL}$ will still remain relatively unconstrained: for a detector the size of ATLAS ($\mathcal{O}(10\text{ m})$), any particle with lifetime $c\tau \gtrsim \text{10\text{ m}}$ will be unlikely to decay inside the detector a significant portion of the time. For $m_a = 10\text{ GeV}$, this corresponds to $\overline{C}_{LL}/\Lambda \sim 4 \times 10^{-6}\text{ TeV}$. Hence, regardless of how strong constraints on $C_{ah}$ can get, any ALP with $c\tau \gtrsim 10\text{ m}$ will evade searches in current colliders.

Many BSM scenarios propose the existence of long-lived particles whose average decay lengths lie well beyond the reaches of current particle detectors. As a result, there have recently been proposals for detectors to search for such long-lived particles such as MATHUSLA~\cite{Chou:2016lxi,Curtin:2018mvb, Lubatti:2019vkf,Alpigiani:2020tva}, CODEX-b~\cite{Aielli:2019ivi}, and ANUBIS~\cite{Bauer:2019vqk}. Here we focus on projected constraints for MATHUSLA from Ref.~\cite{Lubatti:2019vkf}. In the reference, the MATHUSLA detector is taken to be located on the surface of the Earth ($\sim$ 100 meters above ATLAS and 100 meters down the beam pipe, with dimensions $100\times 100\times 20 \text{ m}^3$). An estimate is provided in Ref.~\cite{Curtin:2018mvb} for the projected constraint on the Higgs cross section to long-lived particles:
\begin{align}
(\epsilon\cdot\sigma)^{\text{MATH}} &\approx \frac{4}{\mathcal{L}\,n_{\text{LLP}}\,P_{\text{decay}}^{\text{MATH}}(c\tau)},\label{eq:MATH_constraint}\\
P_{\text{decay}}^{\text{MATH}} &= \epsilon_{\text{geom}}\cdot(e^{-L_2/bc\tau} - e^{-L_1/bc\tau}),
\end{align}
\begin{figure}[t!]
    \centering
    \includegraphics[width=0.8\linewidth]{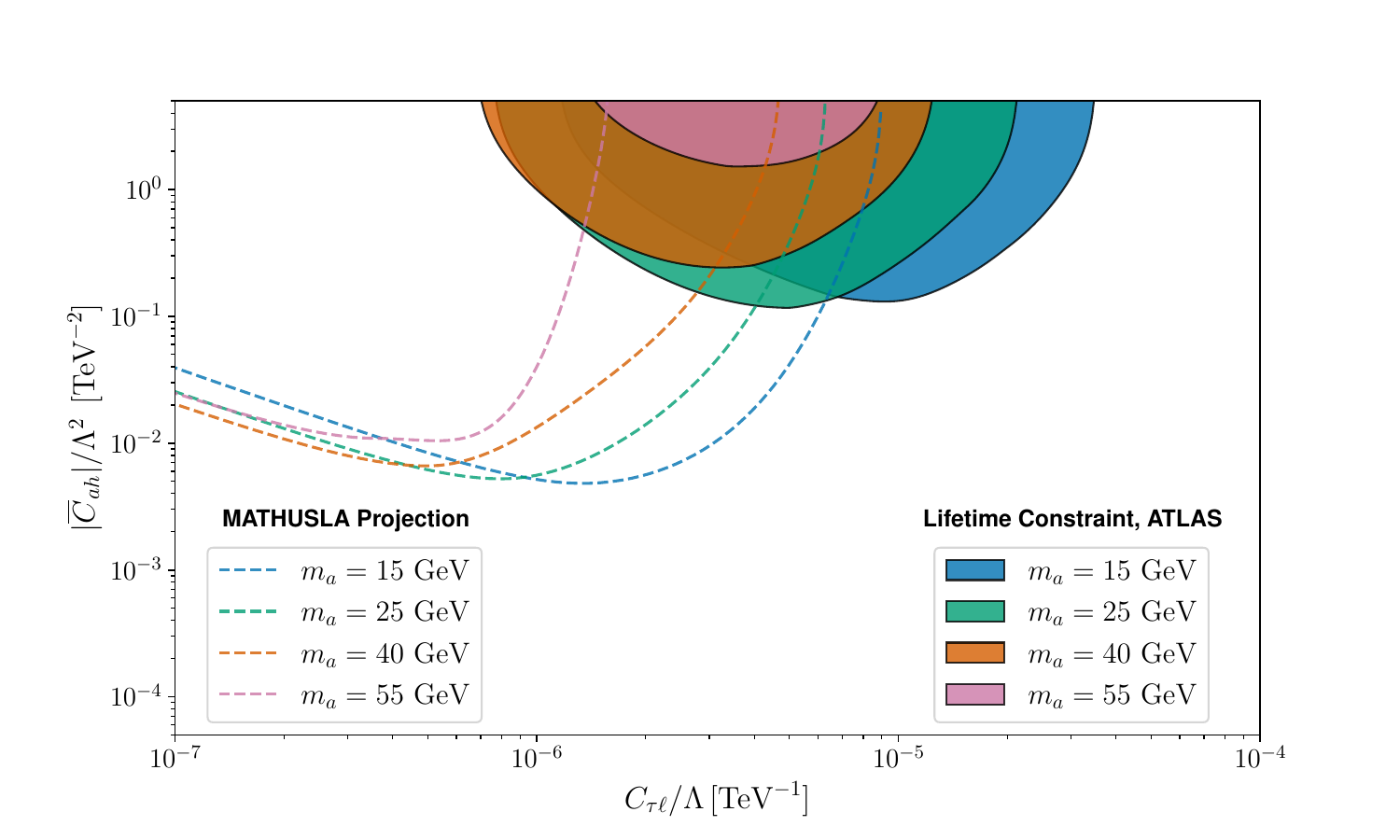}
    \caption{Projected exclusions for the ALP-Higgs coupling from MATHUSLA~\cite{Lubatti:2019vkf} (dashed), with corresponding constraints from Ref.~\cite{Aad:2019xav} (solid regions) for reference. The MATHUSLA results have been updated using the fit described in the text for the masses considered (15, 25, 40, and 55\gev). Note that although in the model we assume a universal lepton coupling $\overline{C}_{LL}$, the exclusion here is essentially on the subset of the couplings $C_{\tau \ell}$.}
    \label{fig:MATH_Cah_vs_Cll}
\end{figure}
where $b = 2m_a/m_h$, $(L_1, L_2) = (200\,\rm{m},\, 230\,\rm{m})$, $\epsilon_{\text{geom}} = 0.05$, and $n_{\text{LLP}}$ is the number of long-lived particles in the Higgs decay channel. We found that this approximation works relatively well for the projected exclusions of $\mathcal{B}(h\rightarrow aa)$ from Ref.~\cite{Lubatti:2019vkf} at long lifetimes, provided we take $(L_1, L_2) = (180\,\rm{m}, 200\,\rm{m})$ (due to the smaller detector size). At lower lifetimes, the approximation fails due to more complicated dependence on the detector acceptance and efficiency. Hence, for low lifetimes, we fit the $\mathcal{B}(h\rightarrow aa)$ projected exclusions from Ref.~\cite{Lubatti:2019vkf} with a polynomial fit, with the condition that it matches Eq.~(\ref{eq:MATH_constraint}) for $c\tau \gtrsim 10\text{ m}$. Details of the fit can once again be found on the Github repository Ref.~\cite{MarcarelliGithub2021}. Our projected constraints on the ALP couplings are shown in Fig.~\ref{fig:MATH_Cah_vs_Cll}; we find tha MATHUSLA can effectively constrain the model for much lower values of $|\overline{C}_{LL}|$.

\subsection{Combined Results}
Here we present the combined results from LFV considerations, prompt decays and displaced decays. To do so, we present exclusion plots on the average lepton coupling $|\overline{C}_{LL}|/\Lambda$ for different Higgs couplings. Previous constraints were placed on $\overline{C}_{ah}$ (see Eq.~(\ref{eq:Cahbar})), so we consider two extremes: $C_{ah} = 0$ and $C_{ah}' = 0$. Both terms $C_{ah}$ and $C_{ah}'$ show up for, e.g., the pion in the Standard Model: $C_{ah}'$ arises from Yukawa couplings with the lighter quarks, whereas $C_{ah}$ arises through a Higgs-gluon effective coupling due to heavy quark loops \cite{Donoghue:1990xh, Grinstein:1988yu}. Since the effect of the $C_{ah}'$ coupling is suppressed by a factor of $m_a^2/m_h^2$, the most relevant Higgs coupling is $C_{ah}$ for light ALP masses. However, in scenarios where one has a heavy composite ALP in a theory with only light quarks, such as the model in Ref.~\cite{Davoudiasl:2017zws}, $C_{ah}'$ is the only significant ALP-Higgs coupling. All constraints are presented at the 95\% CL, except for the MEG LFV constraint, which is at the 90\% CL.

\begin{figure}[h!]
    \centering
    \includegraphics[width=\linewidth]{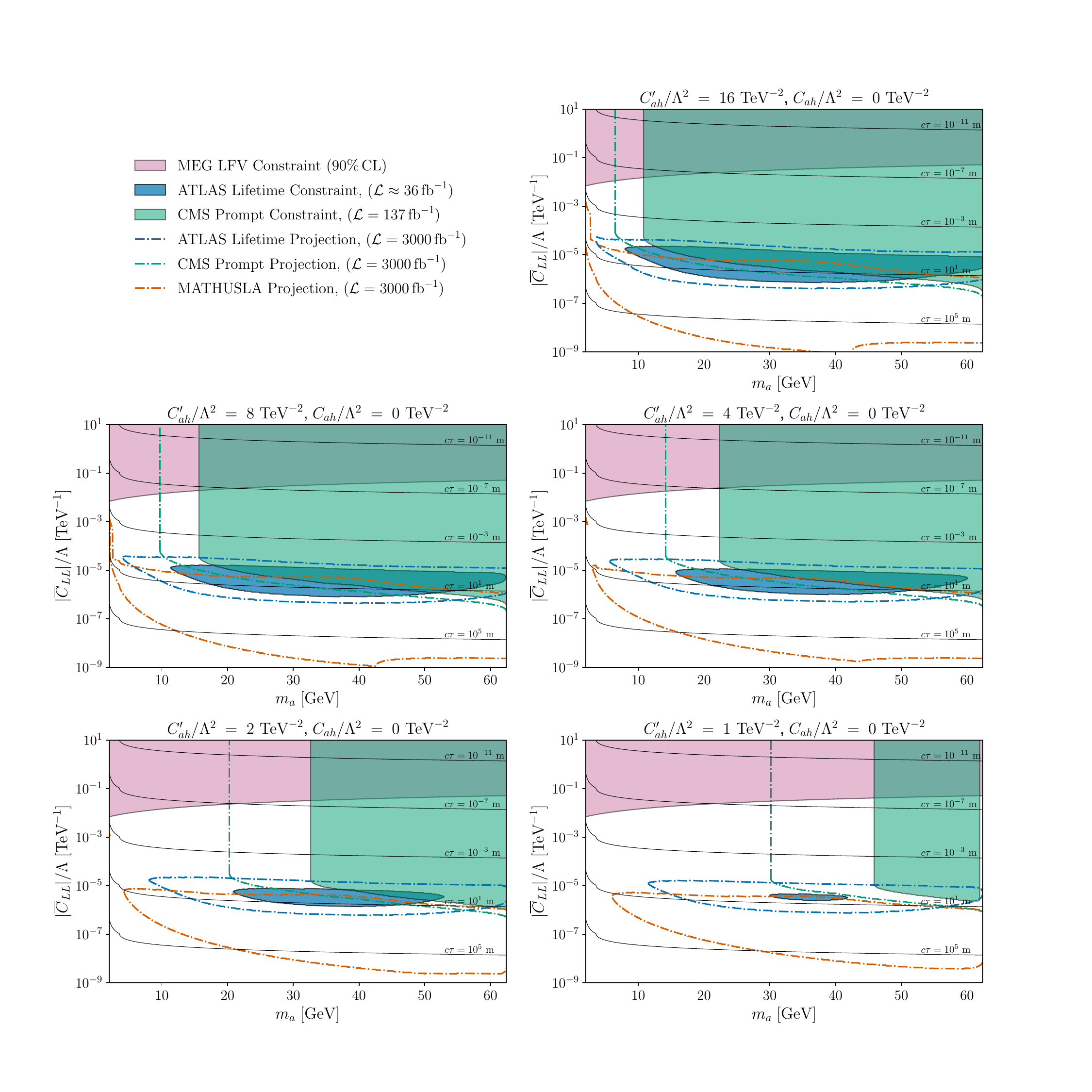}
    \caption{Combined exclusion plots and projections from LFV considerations, prompt decay searches and displaced vertex searches, with $C_{ah} = 0$. As $C_{ah}'$ is decreased, the allowed region of ALP-lepton couplings once again increases dramatically. }
    \label{fig:full_Cahp}
\end{figure}

The resulting plots are demonstrated in Figs.~\ref{fig:full_Cah} and \ref{fig:full_Cahp}. A few observations are in order:
\begin{itemize}
    \item Although these constraints are placed on $\overline{C}_{LL}$ assuming universal on- and off-diagonal lepton couplings, the majority of the exclusions (barring the LFV constraints) mainly constrain $C_{\tau \ell}$. This is because the rely on ALP decays, and the ALP decays into at least one $\tau$ a vast majority of the time. Hence, one could consider a scenario with larger $C_{e\mu}$ coupling for phenomenological purposes without violating these constraints.
    \item In both scenarios, the model is unconstrained for $\overline{C}_{LL}/\Lambda \lesssim 10^{-6}\,\textrm{TeV}^{-1}$, regardless of $C_{ah}^{(\prime)}$.  As a result, there is a lot of unexplored potential for long-lived ALPs, which could potentially be discovered or ruled out by proposed detectors like MATHUSLA.

    \item For low enough $\overline{C}_{ah}$ ($C_{ah}/\Lambda^2\lesssim 0.1\,\textrm{TeV}^{-2}$, $C_{ah}'/\Lambda^2 \lesssim 1\,\textrm{TeV}^{-2}$), the only significant constraints on the ALP-lepton coupling come from LFV considerations. Such constraints can be improved either through increased detector luminosities, or other searches which exploit the lepton flavor violating decays of the ALP.

\begin{figure}[h!]
    \centering
    \includegraphics[width=\linewidth]{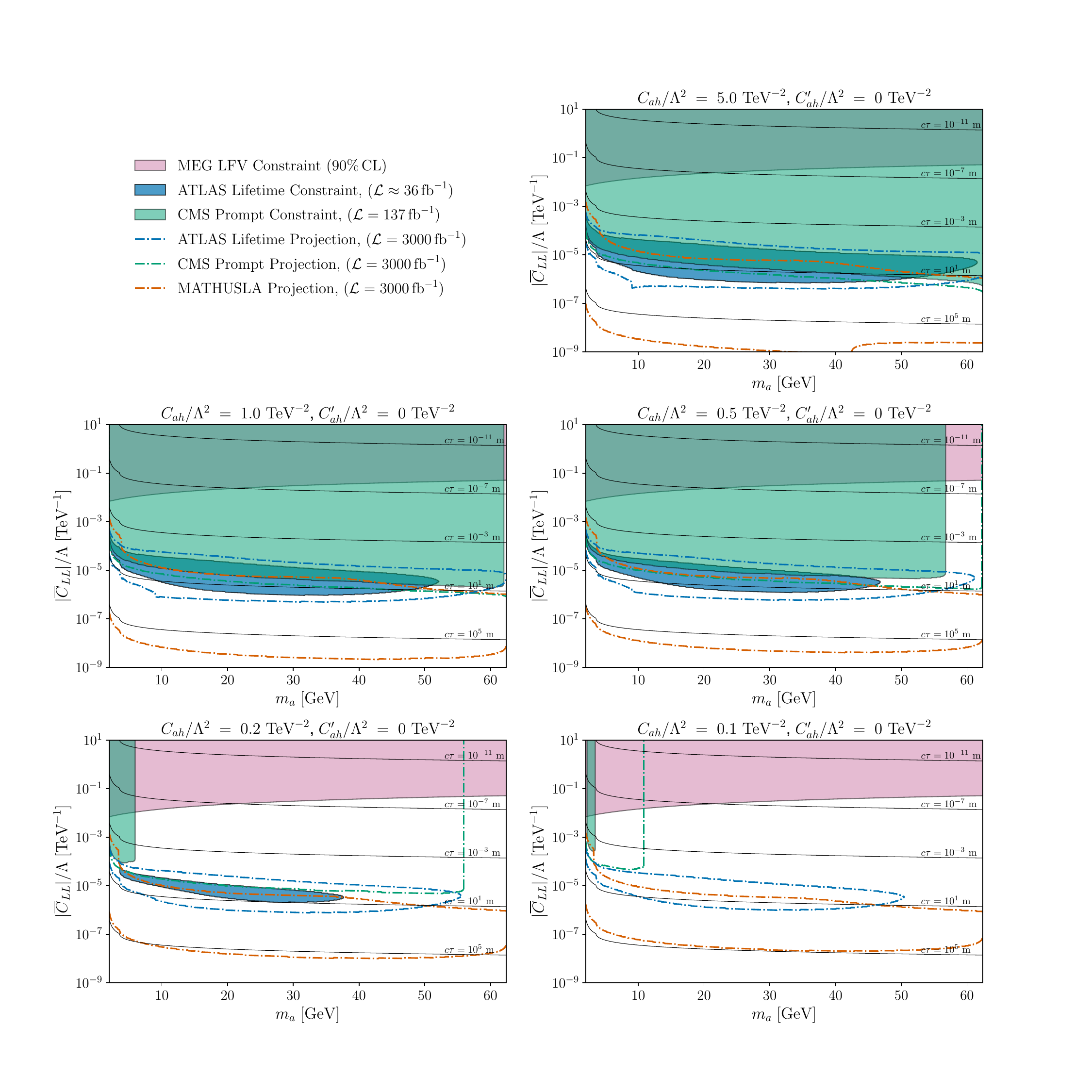}
    \caption{Combined exclusion plots and projections from LFV considerations, prompt decay searches and displaced vertex searches, with $C_{ah}' = 0$. As $C_{ah}$ is decreased, the allowed region of ALP-lepton couplings increases dramatically.}
    \label{fig:full_Cah}
\end{figure}

\end{itemize}

\section{Anomalous Magnetic Moment of the Muon}\label{sec:g-2}
The recent Fermilab measurement \cite{Abi:2021gix} of the muon anomalous magnetic moment $a_\mu=(g_\mu-2)/2$  has confirmed the long-standing Brookhaven result \cite{Bennett:2006fi}.  The combined value of $a_\mu$ from the Fermilab and Brookhaven experiments is at a $4.2\sigma$ disagreement with the SM  predictions as summarized recently by the Muon $g-2$ Theory Initiative \cite{Aoyama:2020ynm}, with $\Delta a_\mu = (25.1\pm5.9)\times 10^{-10}$. In this section, we explore the region of our parameter space which can explain the muon $g-2$ results while still adhering to the LFV constraints from Section \ref{sec:FV}. The region of parameter space necessary enforces an additional hierarchy of diagonal couplings $C_{\ell\ell}$ to off diagonal couplings $C_{\ell\ell'}$, so we recast the constraints accordingly.

The contribution of the anomalous magnetic moment from the LFC and LFV couplings in our model (treating $C_{\gamma\gamma} = 0$) is \cite{Cornella:2019uxs,Bauer:2019gfk,Buen-Abad:2021fwq},
\begin{align}
(\Delta a_\mu)_{\text{LFC}} &= -\frac{m_\mu^2}{4\pi^2\Lambda^2}\left[\frac{\alpha}{\pi} C_{\mu\mu}{\sum_\ell} C_{\ell\ell}\left(H(m_a,m_\mu,m_\ell) + h(m_a,m_\mu,\Lambda)\right) + |C_{\mu\mu}|^2h_1(x_\mu)\right]\label{eq:lfc}\\
(\Delta a_\mu)_{\text{LFV}} &= \frac{m_\mu^2}{16\pi^2\Lambda^2}\left[\left|C_{e\mu}\right|^2h_3(x_\mu) - \frac{m_\tau}{m_\mu}|C_{\mu\tau}|^2g_3(x_\tau)\right]\label{eq:lfv}
\end{align}
where $x_\ell = m_a^2/m_\ell^2$ and the loop functions $h_i(x)$ and $g_3(x)$ are given in Appendix \ref{sec:lfv_fns}, and the functions $H$ and $h$ are two-loop functions from consideration of the loop-induced ALP-photon coupling, defined in Ref.~\cite{Buen-Abad:2021fwq}.  We remind the reader that we have included only parity-conserving couplings in our model; couplings between the axion and lepton vector currents can give additional contributions to $\Delta a_\mu$, see Refs.~\cite{Cornella:2019uxs,Bauer:2019gfk}

The possibility of explaining the anomaly solely with LFV couplings (Eq. (\ref{eq:lfv})) is explored in Ref.~\cite{Bauer:2019gfk}, but without parity-violating couplings much lighter ALP masses than we consider here are required. Explanation of the anomaly with LFV couplings is also constrained by muonium-antimuonium oscillation measurements and will be further tested by future measurements from Belle-II \cite{Endo:2020mev,Iguro:2020rby}. The anomaly was also addressed, using larger ALP masses, more recently in Ref.~\cite{Buen-Abad:2021fwq}, but in their scenario the only non-zero ALP-lepton coupling is $C_{\mu\mu}$.  Here, we consider the scenario in which there exists some hierarchy between the flavor on- and off-diagonal couplings, and negligible tree-level photon coupling, so that the ALP-photon contribution to $\Delta a_\mu$ is induced via lepton loops.

\begin{figure}
    \centering
    \includegraphics[width=0.8\linewidth]{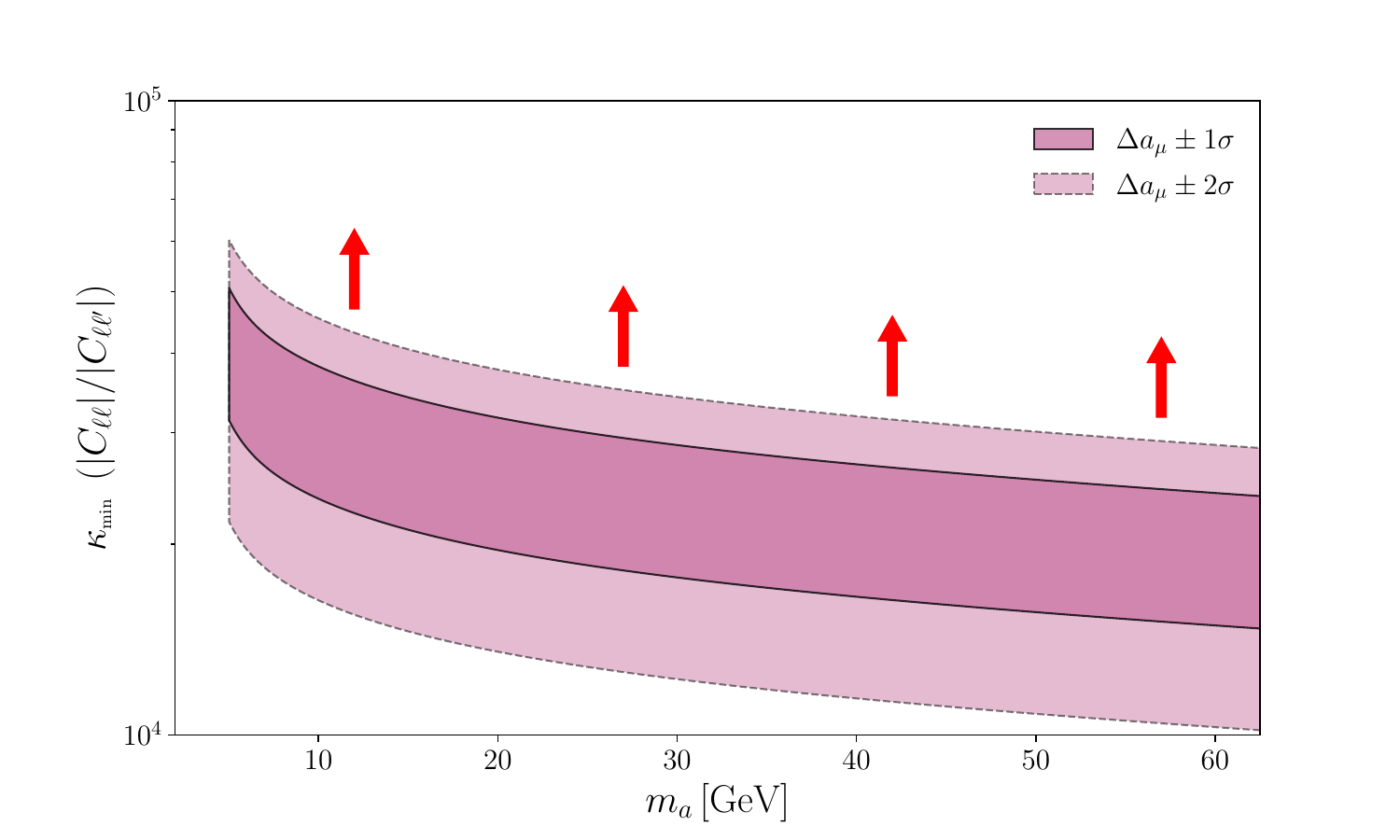}
    \caption{Minimum hierarchy between on- and off- diagonal couplings required to explain $\Delta a_{\mu}$ at the $1\sigma$ and $2\sigma$ level without violating LFV constraints from $\mu \rightarrow e\gamma$. The arrows indicate that this is the {\it minimum} value of the hierarchy, dependent on the true value of $\Delta a_{\mu}$.}
    \label{fig:hierarchy}
\end{figure}

Both $(\Delta a_\mu)_{\text{LFC}}$ and $(\Delta a_\mu)_{\text{LFV}}$ can have the right sign to explain $\Delta a_\mu$ under certain conditions. $(\Delta a_\mu)_\text{LFC} > 0$ if the first term is negative and greater in magnitude than the second, which is generally true if $C_{\mu\mu} < 0$ and $C_{ee},C_{\tau\tau} > 0$. Alternatively, $(\Delta a_\mu)_{\text{LFV}} > 0$ if $|C_{e\mu}| \gtrsim (m_\tau/m_\mu)^{3/2}|C_{\mu\tau}| \sim 70 |C_{\mu\tau}|$. Although both of these are technically possibilities, it turns out that in order to adhere to LFV constraints while explaining the $\Delta a_\mu$ anomaly, we find that the only viable solutions have $|C_{\ell\ell'}| \ll |C_{\ell\ell}|$, so the LFC contribution will be much larger than the LFV contribution. We assume $C_{\mu\mu} < 0$ and $ C_{ee} = C_{\tau\tau} = -C_{\mu \mu}$, and we impose a hierarchy $|C_{\ell\ell}| = \kappa |C_{\ell\ell'}|$.  The dominant contribution to the muon $g-2$ anomaly with these assumptions is from the Barr-Zee diagram \cite{Bjorken:1977vt,Barr:1990vd}.  Through very general considerations, we find that we must have $\kappa \gtrsim 5\times10^4\,\rm{GeV}$; otherwise, $|C_{\mu e}|$ is large enough to violate the constraints from MEG (see Section \ref{sec:FV}). A plot of the minimum hierarchy $\kappa$ necessary to explain the muon $g-2$ discrepancy while adhering to these LFV constraints is shown in Fig.~\ref{fig:hierarchy}. Note that all values above the minimum value $\kappa$ are allowed, including $\kappa\rightarrow \infty$, so that the off-diagonal coupling is taken to 0, but this minimum value is dependent on the true value of $\Delta a_\mu$. This is indicated in the Fig.~\ref{fig:hierarchy} with upward red arrows.

\begin{figure}
    \centering
    \includegraphics[width=0.8\linewidth]{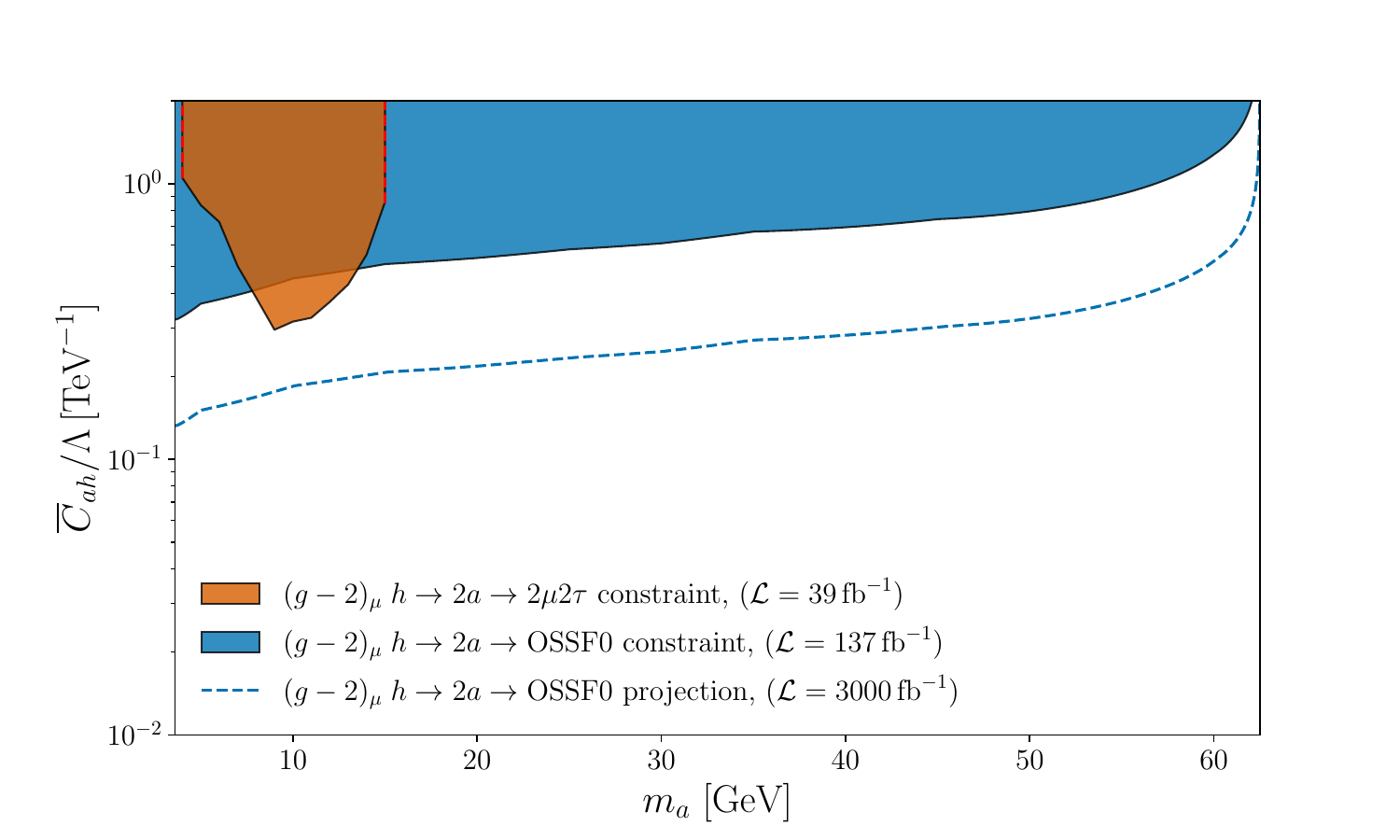}
    \caption{Constraints on $\overline{C}_{ah}/\Lambda$ assuming a coupling hierarchy such that $|C_{\ell\ell'}|\ll |C_{\ell\ell}|$ (in this figure we have taken $\kappa = 5\times 10^4$, but it would look identical for larger $\kappa$). A large hierarchy like this is required to explain the $\Delta a_{\mu}$ discrepancy without violating existing LFV constraints from MEG (see Fig.~\ref{fig:hierarchy}). The main difference between these bounds and the bounds in Fig.~\ref{fig:prompt_Cah_vs_ma} is the hierarchy between on- and off-diagonal couplings, greatly suppressing the $a\rightarrow \tau \mu$ and $a\rightarrow \tau e$ channels.}
    \label{fig:lfv_g-2}
\end{figure}

It is interesting to consider the effect that such a hierarchy has on our phenomenology as well.  The main difference in the phenomenology is that previous works on pseudoscalar decays of the Higgs with diagonal lepton couplings such as \cite{Sirunyan:2020eum,Sirunyan:2019gou}  become more relevant. We find that for some range of the masses considered in \cite{Sirunyan:2019gou} ($4\,\textrm{GeV}\leq m_a \leq 15\,\textrm{GeV}$), their study outperforms the OSSF0 analysis from \cite{Sirunyan:2019bgz} for the hierarchical case. Although the OSSF0 search outperforms the $h \rightarrow 2a\rightarrow 2\mu 2\tau$  search for most masses considered, the latter was conducted with less than a third the luminosity ($39\,\rm{fb}^{-1}$ vs. $137\,\rm{fb}^{-1}$). With increased luminosity one can certainly expect better results, although we do not attempt to estimate such a projection here.

Explicitly solving for the $C_{\ell\ell}$ required to explain $\Delta a_\mu$ in Eq.~\ref{eq:lfc} (again, assuming $C_{ee} = -C_{\mu\mu} = C_{\tau\tau} > 0$) yields $|C_{\ell\ell}| / \Lambda \approx 15\,\rm{TeV}^{-1}$ over the range of masses considered. As a result of this large flavor-conserving lepton coupling, the value of $|C_{\gamma \gamma}^{\rm eff}|$ is necessarily much larger in this scenario (see Eq.(\ref{eq:Cggeff})), so we must also reconsider experimental constraints in which the ALP-photon coupling is probed directly.  As summarized in Ref.~\cite{Agrawal:2021dbo}, the strongest current constraint for $m_a \gtrsim m_\tau$ comes from ATLAS \cite{Aad:2020cje} and CMS \cite{Sirunyan:2018fhl} searches for light-by-light scattering in Pb-Pb collisions.  Using the narrow-width approximation, the predicted cross-section in our model is
\begin{equation}
\sigma(\gamma \gamma \rightarrow a \rightarrow \gamma \gamma) = \frac{8\pi^2}{m_a} \Gamma(a \rightarrow \gamma \gamma) \mathcal{B}(a \rightarrow \gamma \gamma) I(m_a^2),
\end{equation}
where the function $I(m_a^2)$ captures the energy dependence of the photon-photon luminosity \cite{Baur:1990fx,Knapen:2016moh}.  In the current scenario where the coupling $C_{\gamma \gamma}^{\rm eff}$ arises solely from lepton loops and the ALP satisfies $m_a > m_\tau$, the decay modes of $a$ are still dominantly into pairs of leptons and the constraint from Pb-Pb collisions is not significant.

\section{Conclusions \label{sec:conclusions}}

The central focus of this paper has been on the decay of the Higgs into two pairs of off-diagonal leptons, through ALP mediators.  Multiple theories have suggested the presence of a heavy axion-like particle leading to new physics, but aspects of the combination of the Higgs decay and lepton flavor violation were yet to be fully explored.  Useful constraints on large regions of the ALP mass and coupling constant parameter space can be obtained from various collider searches for both prompt and long-lived particles.

Initially working under the assumption that the coupling of the ALP to leptons $C_{\ell\ell'}$ was universal across all combinations of $\ell$ and $\ell',$ formulae were derived for the first-order decay rates of the Higgs into a pair of ALPs and the ALP into a pair of leptons.  The process $h\to aa \to $ OSSF0 was then simulated with MadGraph which, when combined with cuts from ATLAS search \cite{Sirunyan:2019bgz}, made it possible to immediately place bounds on $\overline{C}_{ah}$ and $m_a$ (Fig.~\ref{fig:prompt_Cah_vs_ma}). There is a possibility that other future searches can lead to better constraints by taking advantage of the LFV decay modes. In particular, Ref.~\cite{Sirunyan:2019gou} considers constraints on the branching fraction of $h\rightarrow 2a \rightarrow 2\mu 2\tau$ and $h\rightarrow 2a\rightarrow 4\tau$ by examining the invariant mass distribution of $a$ candidates. The addition of $\mathcal{O}(1)$ LFV couplings could lead to many more events in the $h\rightarrow 2a\rightarrow 2\mu 2\tau$ mode via $a\rightarrow \tau \mu$, which would likely affect the invariant mass distribution. It would be interesting to see if constraints produced in this manner would outperform those obtained from the OSSF0 signal, especially for larger masses, where the constraints from \cite{Sirunyan:2019gou} appear to drop off.

These bounds were combined with others from existing investigations into lepton conversion processes such as $\mu\to e\gamma$ and $\tau\to \ell\gamma$, by adapting the formulas for the rates of those processes from Ref.~\cite{Cornella:2019uxs} to our $C_{\ell \ell'}$ and $C_{\gamma\gamma}^\text{eff}$ and plotting the resulting constraints (Fig.~\ref{fig:full_Cahp}). Of these, the $\mu\to e\gamma$ bound obtained from the data of the MEG experiment was found to be the most powerful.

Next, we investigated the possibility that the ALP vertices from the Higgs decay process might be displaced.  After charting where in the parameter space such events would be allowed by the dimensions of the detector (Fig.~\ref{fig:prompt_Cah_vs_ma_lifetimes}), we derived using data from Ref.~\cite{Aad:2019xav} further bounds on the relationship between $\overline{C}_{ah}$ and $C_{\ell\ell’}$ (Fig.~\ref{fig:lifetime_prompt_Cah_vs_Cll}).  These bounds successfully constrain lower values of the parameters than the prompt decays,  by a factor of about 3.  The model is still unconstrained however, for $|\overline{C}_{ah}|/\Lambda^2 \lesssim 0.1\text{ TeV}^{-2}$, $C_{\ell\ell'} /\Lambda \lesssim 10^{-6}\ \text{TeV}^{-1}$, or for ALPs with lifetimes exceeding around 10 m.   In light of the limitation on the lifetimes in particular, the advancements to these constraints that a larger detector like MATHUSLA could produce were examined.  The projected bounds from MATHUSLA were added to our displaced vertex plots, where it was found that the new detector could make the constraints multiple orders of magnitude more stringent in some cases (Fig.~\ref{fig:MATH_Cah_vs_Cll}).


Finally, we investigated adapting our model to explain the recent results for the muon $g-2$ anomaly. Because of the ALP's large mass and because it is a pseudoscalar, the largest contribution to the process must come from the Barr-Zee diagram for the sign of the difference to be correct.  If one allows for a hierarchy of at least four orders of magnitude between flavor diagonal and off-diagonal values of $C_{\ell\ell'}$, the desired contribution can be achieved.  This is encouraging, as it implies that with some adjustment, the OSSF0 signal could be another way to search for a deeper explanation for this anomaly.

Our analysis so far has been restricted to $m_a \gtrsim 2$ GeV, so that decay modes involving at least one $\tau$ lepton are dominant due to the mass scaling of the leptonic ALP decay rate.  For ALPs with masses below the $\tau$ mass, the only possible on-shell decay mode would be $a \rightarrow \mu e$, which could lead to the signal $h \rightarrow (\mu e) (\mu e)$.  It would be interesting to study a dedicated search in this channel in future work; the signal with displaced decays in particular could be very distinctive.  Study of the more general parity-violating case where couplings of the ALP to the lepton vector currents are nonzero would also be an interesting extension of this work.

\section*{Acknowledgements}
We thank Keith Ulmer for helpful conversations about this work.  We thank Claudia Cornella, Paride Paradisi and Olcyr Sumensari for useful correspondence about limits due to lepton flavor violation searches.  We also thank Manuel Buen-Abad, JiJi Fan, Matt Reece, and Chen Sun for helpful correspondence about axion contributions to muon $g-2$ and constraints on photon-photon interactions.  This work is supported by the U.S. Department of Energy under Grant Contracts DE-SC0012704 (H.~D.) and DE-SC0010005 (E.~N. and R.~M.).

\begin{raggedright}
\bibliography{comp_n_dm}
\end{raggedright}

\appendix

\section{Lepton Flavor Violation Constraints\label{sec:lfv_fns}}
The form factor in Eq.~(\ref{eq:leptconv}) is given by 
$\mathcal{F}_2^{\ell\ell^\prime}(0) = \mathcal{F}_2^{\ell\ell'}(0)_{\text{lin}.} + \mathcal{F}_2^{\ell\ell'}(0)_{\text{quad.}}$, where
\begin{align}
    \mathcal{F}_2^{\ell\ell'}(0)_{\text{lin.}} = -&e^2\frac{m_{\ell}^2}{8\pi^2\Lambda^2}C_{\ell\ell^\prime}C_{\gamma\gamma}^{\rm{eff}}g_{\gamma}(x_{\ell}),\label{eq:F2lin}\\
    \mathcal{F}_2^{\ell\ell'}(0)_{\text{quad.}} = - &\frac{m_{\ell}}{16\pi^2\Lambda^2}C_{\ell\ell^\prime}\left[C_{\ell\ell}m_{\ell}g_1(x_\ell) + C_{\ell^\prime\ell^\prime}m_{\ell^\prime}g_2(x_\ell)\right]\nonumber\\
    \ \ -&\frac{m_\mu m_\tau}{32\pi^2\Lambda^2}\begin{cases}
    C_{\tau e}C_{\tau\mu}g_3(x_\tau), & (\mu \rightarrow e\gamma)\\
    C_{\tau \mu}C_{\mu e}g_4(x_\tau), & (\tau \rightarrow e\gamma)\\
    C_{\tau e}C_{\mu e}g_4(x_\tau), & (\tau \rightarrow \mu\gamma)
    \end{cases}
\end{align}
and $x_\ell = m_a^2/m_{\ell}^2 - i\epsilon$ with $\epsilon\rightarrow 0$. The loop functions are given by Ref.~\cite{Cornella:2019uxs}
\begin{align}
    g_\gamma(x) &= 2\log{\frac{\Lambda^2}{m_a^2}} - \frac{\log{x}}{x-1} - (x-1)\log\left(\frac{x}{x-1}\right) - 2,\\
    g_1(x) &= \frac{x-3}{x-1}x^2\log{x} + 1 - 2x - 2x\sqrt{x^2-4x}\log{\left(\frac{\sqrt{x}+\sqrt{x-4}}{2}\right)},\\
    g_2(x) &= 1 - 2x + 2(x-1)x\log{\left(\frac{x}{x-1}\right)},\\
    g_3(x) &= \frac{2x^2\log{x}}{(x-1)^3} + \frac{1-3x}{(x-1)^2},\\
    g_4(x) &= 1 - 2x + 2(x-1)x\log\left(\frac{x}{x-1}\right),\\
    h_1(x) &= 1 + 2x - (x-1)x\log{x} + 2x(x-3)\frac{\sqrt{x(x-4)}}{x-4}\log\left(\frac{\sqrt{x}+\sqrt{x-4}}{2}\right)\\
    h_2(x) &= 1 + \frac{x^2}{6}\log{x} - \frac{x}{3} - \frac{x+2}{3}\sqrt{(x-4)x}\log\left(\frac{\sqrt{x}+\sqrt{x-4}}{2}\right)\\
    h_3(x) &= 2x^2 \log\frac{x}{x-1} - 1 - 2x
\end{align}

\pagebreak

\end{document}